\documentclass[12pt,english]{article}

\usepackage[T1]{fontenc}
\usepackage[utf8]{inputenc}
\usepackage{mathpazo} 

\usepackage{geometry}
\usepackage{setspace}
\usepackage[english]{babel}

\usepackage{amsmath,amsthm,amssymb,amstext}

\usepackage{graphicx}
\usepackage{float}
\usepackage{booktabs}
\usepackage{array,makecell}
\usepackage[flushleft]{threeparttable}
\usepackage{longtable}
\usepackage{rotating}
\usepackage{subfig}

\usepackage[font=small]{caption}

\usepackage{tikz}
\usepackage{pgfplots}

\usepackage{titlesec}
\titlespacing*{\paragraph}{0pt}{1.5ex plus 1ex minus .2ex}{1em}

\usepackage[
  unicode=true,
  pdfusetitle,
  bookmarks=true,
  bookmarksnumbered=false,
  bookmarksopen=false,
  breaklinks=false,
  pdfborder={0 0 1},
  backref=false,
  colorlinks=false 
]{hyperref}

\hypersetup{
  colorlinks=true,
  citecolor=blue,
  linkcolor=blue,
  urlcolor=blue
}

\newtheorem{theorem}{Theorem}

\newtheorem{lemma}[theorem]{Lemma}
\newtheorem{proposition}{Proposition}

\usepackage[comma,authoryear,round]{natbib}

\usepackage[toc,page,header]{appendix}
\usepackage{minitoc}

\usepackage{eurosym}
\usepackage{accsupp}
\DeclareRobustCommand{\officialeuro}{%
  \BeginAccSupp{method=hex,unicode,ActualText=20AC}%
    \ifmmode\expandafter\text\fi{\fontencoding{U}\fontfamily{eurosym}\selectfont e}%
  \EndAccSupp{}%
}

\usepackage{datetime}
\newdateformat{monthyearformat}{\monthname[\THEMONTH] \space \THEYEAR}

\title{The sustainability of contribution norms with income dynamics  }

\author{Pau Juan-Bartroli\footnote{This manuscript was previously distributed as \textit{Social Mobility and Long-Run Cooperation}. For comments and discussions in either version of the manuscript, we thank Ingela Alger, Alae Baha, Astrid Hopfensitz, Gerard Maideu-Morera, Sébastien Pouget, Jaume Ventura, and Takuro Yamashita. We acknowledge funding from the European Research Council (ERC) under the European Union's Horizon 2020 research and innovation programme (grant agreement No 789111 - ERC EvolvingEconomics).}  \\
	\textit{Toulouse School of Economics}
	\and 
	Esteban Muñoz-Sobrado \\
	\textit{Universitat Rovira i Virgili}} 
\date{\monthyearformat\today}

\begin{document}

\maketitle

\begin{abstract}
The sustainability of cooperation is crucial for understanding the progress of societies. We study a repeated game in which individuals decide the share of their income to transfer to other group members. A central feature of our model is that individuals may, with some probability, switch incomes across periods—our measure of income mobility—while the overall income distribution remains constant over time. We analyze how income mobility and income inequality affect the sustainability of \textit{\textit{contribution norm}s}—informal agreements about how much each member should transfer to the group. We find that greater income mobility facilitates cooperation. In contrast, the effect of inequality is ambiguous and depends on the progressivity of the \textit{contribution norm} and the degree of mobility. We apply our framework to an optimal taxation problem to examine the interaction between public and private redistribution. 
\end{abstract}
\newpage
\section{Introduction}

Income inequality and income mobility jointly shape incentives for cooperation.\footnote{Given that we keep the income distribution constant, income mobility is interpreted as \textit{relative income mobility}: changes in the relative positions of individuals within the income distribution over time \citep{shorrocks1978income}. This is particularly relevant in the intragenerational context, where mobility is viewed as movement within a fixed income structure rather than across generations \citep{jantti2015income}.} While inequality provides a snapshot of the income distribution, mobility reflects the potential for individuals to change their income over time. Milton Friedman famously noted that the importance of inequality depends on whether it reflects temporary differences or entrenched long-run status: “\emph{A major problem in interpreting evidence on the distribution of income is the need to distinguish between two basically different kinds of income inequality: temporary, short-run differences in income, and differences in long-run income status}” \citep[p. 171]{friedman1962capitalism}. 
This quote emphasizes the instrumental role of mobility in shaping the normative interpretation of inequality.
Despite the importance of income mobility, it is rarely incorporated into theoretical models of cooperation. This omission can lead to misleading predictions: the same level of inequality may foster or hinder cooperation depending on how likely individuals are to change their income over time. \par

This paper examines how the interaction between inequality and mobility affects cooperation in repeated interactions. To investigate this interaction, we develop a tractable theoretical framework that captures the interplay between income inequality, income mobility, and \textit{contribution norms}. The model features three key parameters: a mobility parameter that governs the likelihood of individuals changing their income over time, an inequality parameter that determines the dispersion between individuals’ incomes, and a progressivity parameter that characterizes the shape of the \textit{contribution norm}—how society dictates transfers should vary with income. This parsimonious structure allows us to isolate each component's effect and analyze how their interaction influences the sustainability of cooperation. While previous work has emphasized that income mobility is instrumentally valuable because it reduces long-term inequality \citep{jantti2015income}, our contribution is to show that mobility also plays a critical role in sustaining cooperation.

\paragraph{Setting}

The simplest version of the model consists of an organization with two individuals who differ in their incomes. They play an \textit{infinitely repeated stochastic game} where, in each period, they decide what share of their income to transfer to the organization, which equally distributes transfers among its members. This stage game is infinitely repeated, and individuals maximize the discounted sum of their utility, given complete information about the history of contributions. Contributing to the organization is individually costly—any amount transferred reduces one’s material payoff—but aggregate utility is maximized when both individuals transfer their entire income to the organization. 
Although individuals are solely concerned with their material payoffs, even high-income individuals may have incentives to contribute to the organization, as their prospect of becoming low-income in the future makes the benefits of redistribution valuable to them over time.

Our model has three key features. First, with some probability, individuals switch roles: the rich individual becomes poor, and the poor individual becomes rich. We model income mobility as a homogeneous Markov process over the set of individual rankings, where each state represents a specific ordering of individuals by income. This transition is governed by a parameter $m \in [0,1]$, which captures the degree of income mobility within the organization. When $m = 0$, the individuals remain in the same role in all periods; the organization has no income mobility. When $m = 1$, the individuals have the same probability of moving to either role; the organization has full income mobility. More generally, higher values of $m$ correspond to a greater probability of switching roles and, thus, greater income mobility within the organization. This approach aligns with standard stochastic models of income dynamics while allowing us to embed mobility into a repeated-game framework where strategic cooperation is endogenous.



Second, the organization’s income distribution remains constant over time. Although individuals may change their income level, the organization's total income and income levels are fixed. 
To measure income inequality, we use the \textit{Atkinson index} in which a parameter $\alpha \geq 0$ governs income inequality (\citealp{ATKINSON1970244}). When $\alpha = 0$, all income types receive the same share of the endowment; the organization has full income equality. When $\alpha \to \infty$, all income becomes concentrated in the highest income type; the organization has full income inequality. More generally, the higher $\alpha$, the higher the inequality within the organization. \par 
Third, the organization's members have an informal shared agreement on the transfers individuals should make to the organization given their income, referred to as the \textit{contribution norm} (\citealp{reuben2013enforcement}). This agreement can be interpreted as a social norm of behavior that dictates how people should behave in society (\citealp{bicchieri2008fragility}; \citealp{burke2011social}). We consider a set of norms parametrized by a parameter $\beta \geq 0$ that governs the progressivity of the agreement. When $\beta \in (0, 1)$, the rule is \textit{regressive}: rich individuals contribute more in absolute terms but less in relative terms. When $\beta > 1$, the rule is \textit{progressive}: rich individuals contribute more in absolute and relative terms. \par
In this paper, instead of focusing on characterizing the set of Subgame Perfect Nash Equilibria (SPNE), we focus on characterizing the lowest discount factor $\underline{\delta}$ such that the outcome in which all individuals follow the \textit{contribution norm} is an SPNE. As is standard in repeated games, our analysis concentrates on the efficient cooperative equilibrium among the set of sustainable outcomes, rather than characterizing the entire equilibrium set.

In our context this can also be justified for three reasons. First, this outcome is normatively appealing: it represents the outcome in which all individuals follow the group’s agreement, aligning with the government's perspective or the community norms (\citealp{caplin2008foundations}; \citealp{dold2018toward}).
Second, cooperative outcomes are commonly observed in empirical settings involving informal redistribution, such as risk-sharing arrangements in villages or among co-workers (\citealp{dercon2005risk}; \citealp{dubois2008formal}; \citealp{mobarak2013informal}).
Third, focusing on a unique outcome allows for transparent comparative static analysis, which would be challenging with the set of SPNE.

\paragraph{Results} We show three main results. First, as long as the organization has a positive degree of income mobility, the \textit{contribution norm} can be sustained if individuals are sufficiently patient. This result is important as cooperation is never sustainable when $m = 0$. Hence, cooperation is discontinuous when changing from zero to positive mobility. 
More generally, income mobility facilitates cooperation among its members for any level of income inequality and progressivity of the \textit{contribution norm}.\footnote{More formally, the lowest discount factor $\underline{\delta}$ such that the cooperative outcome is an SPNE is decreasing on $m$ for any $\alpha > 0$ and $\beta \geq 0$.} Intuitively, income mobility enhances high-income individuals’ incentives to cooperate, as it raises the probability that they might eventually become low-income, benefiting from the \textit{contribution norm}. This prospect strengthens the value of cooperation as a form of intertemporal insurance, encouraging individuals to contribute even when they are currently better off.

\par 
Second, inequality has an ambiguous effect on cooperation, which crucially depends on the degree of progressivity of the \textit{contribution norm}. When \textit{contribution norms} are regressive, higher inequality has a positive effect on cooperation. On the other hand, when \textit{contribution norms} are progressive, the relationship between inequality and cooperation is non-monotonic, especially when income mobility is low. Intuitively, when \textit{contribution norms} are regressive, higher inequality makes it less costly for the rich to comply, strengthening cooperation. In contrast, with progressive norms, increasing inequality places a greater burden on the rich, which can discourage compliance, especially when the rich expect to remain in that position in the future.

\par 

Finally, we examine which \textit{contribution norms} are most conducive to sustaining cooperation in the long run. For each level of inequality and mobility, we identify the rule that makes cooperation easiest to sustain, that is, the one requiring the least patience from individuals. This provides an efficiency benchmark that summarizes how the optimal degree of progressivity varies with inequality and mobility.\footnote{This efficiency criterion can also be interpreted as the outcome of a long-run evolutionary process in which norms that better sustain cooperation are more likely to persist or be imitated; see \citet{witt1993evolutionary, safarzynska2010evolutionary, alger2007family, alger2010kinship, alger2025norms}.}


We characterize how the optimal norm depends on inequality and mobility. A central finding is that the selected norm can be either progressive or regressive, depending on the environment.
More generally, as inequality rises, the optimal norm initially becomes more progressive but eventually becomes less progressive beyond a critical level of inequality. In contrast, higher mobility consistently reduces the optimal degree of progressivity, as individuals are more likely to move across income ranks.


\paragraph{Application}
We extend the model to study the interaction between private and public redistribution (\citealp{alesina2005fairness}; \citealp{krueger2011public}). Specifically, we add a policy stage in which a benevolent utilitarian government sets a proportional income tax before individuals engage in voluntary transfers. The government collects taxes and redistributes a fraction of the revenue equally among group members. After taxation, individuals decide how much of their disposable income to voluntarily transfer to others. We analyze how the optimal tax rate depends on income inequality and mobility, and how it interacts with the prevailing \textit{contribution norms}.

The analysis establishes a regime-dependent relationship between public and private redistribution. When voluntary cooperation cannot be sustained for any tax rate, the government is the sole provider of redistribution, and the optimal tax rate increases with income inequality. When voluntary cooperation is sustainable, voluntary transfers reduce the need for public redistribution. In this case, the welfare-maximising tax rate decreases with inequality, income mobility, and the progressivity of the \textit{contribution norm}. These results indicate that public and private redistribution are substitutes and that the design of tax policy depends on the incentive compatibility of private cooperation. Ignoring endogenous cooperation can lead to different policy prescriptions, especially in settings with high mobility or well-defined informal norms. This complements previous work on the political economy of redistribution, which typically assumes either public or private transfers as exogenous \citep[e.g.,][]{alesina2005fairness, benabou2001social}.

\paragraph{Related Literature}

This paper relates to several strands of literature on income mobility, redistribution, and informal cooperation. Empirical studies document substantial variation in income mobility across space and time \citep{chetty2014united, chetty2017fading, deutscher2023measuring}. In the United States, relative mobility has remained stable, but rising inequality has magnified the consequences of the birth lottery \citep{chetty2014united}.
Different mobility metrics can also yield different policy implications \citep{deutscher2023measuring}, underscoring the importance of mobility as both an outcome and a determinant of expectations.\footnote{For a broader review of the link between inequality and mobility, see \citet{durlauf2022great}. For comprehensive overviews of mobility concepts, measurement, and trends, see \citet{fields1999measuring} and \citet{jantti2015income}. 
}

Theoretically, models such as \citet{benabou2001social} and \citet{acemoglu2018social} formalize mobility using Markovian income and status transitions, respectively. Our framework differs from these studies by focusing on intra-personal (rather than intergenerational or group-based) mobility, allowing for downward income transitions and abstracting from dynastic preferences. We examine how such individual income dynamics affect the sustainability of redistribution through voluntary transfers, holding group structure and intergenerational links constant.

Our analysis also builds on the literature on informal risk-sharing under limited commitment \citep{thomas1988self, kocherlakota1996implications}. In these settings, cooperation is sustained by future punishment threats, and efficient allocations are constrained by enforceability. We depart from this work by considering an environment with $N \geq 2$ individuals and introducing income mobility as a central parameter. This allows us to isolate how dynamic risk exposure interacts with cooperation incentives in group settings in which individuals may use social norms to coordinate and insure themselves against negative income shocks.

%

\paragraph{Outline} The remainder of the paper is organized as follows. In Section \ref{Theoreticalmodel}, we present the theoretical framework. In Section \ref{GeneralSol}, we characterize the general solution. In Section \ref{Comparative}, we conduct several comparative statics. In Section \ref{GovReds}, we apply our framework to an optimal taxation problem. In Section \ref{Conclusions}, we conclude. All mathematical proofs are in Appendix \ref{AppA}.
\section{Theoretical model} \label{Theoreticalmodel}
\subsection{Stochastic game} \label{Stochasticgame}
To model repeated cooperation, we consider an \textit{infinitely repeated stochastic game}; a generalization of infinitely repeated games.\footnote{Stochastic games were introduced in \cite{shapley1953stochastic}. For detailed discussions, see \cite{neyman2003stochastic}, \cite{amir2003stochastic}, and \cite{solan2015stochastic}. Stochastic games have been used mainly in resource-extraction problems (\Citealp{levhari1980great}) but also in industrial organization (\Citealp{ericson1995markov}) and inspection problems (\Citealp{baston1991generalized}; \Citealp{darlington2023stochastic}).} In stochastic games the stage game may change over time due to players' behavior and chance. In our setting, we consider a stochastic game with (i) a finite set of states, (ii) a finite number of players, (iii) a common action set, (iv) a transition probability matrix that only depends on chance, and (v) perfect monitoring. More formally, the game is a tuple $(S, N, A, P, \Pi)$, where:
\begin{itemize}
    \item $S$ is a finite set of states.
    \item $N$ is a finite number of players.
    \item $A = A_1 \times ... \times A_n$ is the action set.
    \item $P: S \times S \rightarrow [0, 1]$ is a transition probability function. 
    \item $\Pi = \pi_1 \times \pi_2 \times ...\times \pi_n$ is the payoff function set.
\end{itemize}
In our application, the states represent the income rankings of the players, while the transition probability function $P$ represents the probability of moving to any state at $t + 1$ given the state in period $t$. \par 
\subsection{Stage game} \label{Stagegame}
Each stage game is composed of $N \geq 2$ players who differ in which of the $N$ positions they have been assigned. Each position is allocated a deterministic income $w_1 > ... > w_N > 0$. Thus, the player assigned to position $i$ receives an income of $w_i$. There is a total of $N!$ states, each representing an income rank in the organization. 
After players learn their positions, they decide the share of their income to transfer to the organization. The total amount transferred is equally shared among the group members.
The amount transferred by player $i \in \{1, ..., N\}$ in position $k \in \{1, ..., N\}$, in period $t \in \{1, 2, ...\}$ is denoted by ${x_{i, k}^t} \in [0, 1]$.  \par
Players' material payoff in the stage game depends on players' choices and their incomes. For simplicity, we assume that in the first period, individual $i$ is allocated to position $i$, and refer to ${x_{i, i}^1}$ as ${x_{i}^1}$. In that case, player $i$'s material payoff at $t = 1$ is given by:
\begin{equation} \label{payoff}
\pi_{i}({x_{i}^1};{x_{-i}^1}) = w_i(1 - x_{i}^1) + \frac{1}{N} \sum_{j = 1}^{N} x_{j}^1 \ w_j, 
\end{equation}
where ${x_{-i}^1}$ refers to the strategy profile of all players except $i$.
Players evaluate their material payoff with an increasing and strictly concave function $u$ (i.e., $u' > 0$ and $u '' < 0$). Thus, players decrease their utility when transferring money to the organization, but having all players doing so is socially efficient.\footnote{This formulation is equivalent to assuming that players evaluate their monetary payoffs with a linear function $u(x) = x$ and that contributions to the organization are multiplied by a parameter $\gamma > 1$, as in linear public goods games (\citealp{zelmer2003linear}).} \par 
\subsection{Income mobility} \label{Incomemobility}

{
We model income mobility as a homogeneous Markov process over the set of states. The transition matrix specifies the probability of moving from one state to another across periods, assuming that transitions are governed by chance and not by individual actions. The model satisfies three standard assumptions from the mobility literature: (i) the \textit{Markov property}---transitions depend only on the current state and not on prior history \citep{mcfarland1970intragenerational}; (ii) \textit{time-homogeneity}---transition probabilities are constant over time \citep{shorrocks1978a}; and (iii) \textit{exogeneity}---mobility arises independently of individuals' behavior, reflecting institutional or environmental randomness \citep{prais1955,fields1999measuring}. \par 
The degree of mobility is governed by a single parameter \( m \in [0,1] \), with \( m = 0 \) representing full persistence (identity matrix) and \( m = 1 \) corresponding to maximal mobility (uniform transitions across permissible states). Our formulation maintains a fixed income distribution across periods and captures pure exchange mobility through probabilistic re-rankings. This structure aligns with the literature on transition matrices while enabling us to embed mobility into a repeated-game framework with strategic interactions. When $N = 2$, we consider the following transition matrix: }

\[
   \bordermatrix{~ & s_{ij} & s_{ji} \cr
                s_{ij} & 1 - \frac{1}{2}m & \frac{1}{2}m \cr
                s_{ji} & \frac{1}{2}m & 1 - \frac{1}{2}m }
\]
where $s_{ij}$ represents the state where player $i$ and $j$ are in the first and second position, respectively.\footnote{The income distribution is constant across periods and states. That is, there is always a player with a high and one with a low income.} The rows of the matrix represent the state at period $t$, and the columns represent the state at $t+1$. Each cell represents the probability of moving from a given state at $t$ to a given state at $t+1$. The parameter $m \in [0, 1]$ represents the \textit{degree of income mobility}. When $m = 0$, the organization has no income mobility (i.e., players remain in the same position with certainty). When $m = 1$, there is full income mobility (i.e., each player moves to any position with equal probability). More generally, the larger $m$, the more likely players are to switch positions.  \par 
\par 
When $N > 2$, we define $S'(s) \subset S$ to be the set of states such that there is no player in the same position as in $s \in S$.\footnote{For example, when $s = s_{ijk}$, $S'({s_{ijk}}) = \{s_{jki}, s_{kij}\}$.} We consider the following transition matrix: 
\begin{equation}
\mu_{s, s'} = 
\begin{cases}
    1 - \frac{N-1}{N}m  \hspace{0.6cm} \text{   if }s = s', & \\[0.cm]
   \frac{1}{N}m  \hspace{1.7cm}\text{                   if  }s' \in S'(s), & \\[0.2cm]
   0 \hspace{2.2cm}\text{  otherwise}. & \\[0.2cm]
\end{cases}
\end{equation} \par 
Here, $\mu_{s, s'}$ denotes the probability of moving from state $s$ at period $t$ to state $s'$ at period $t+1$. The transition probability matrix is constructed then by computing $\mu_{s, s'}$ for each pair of states.  
With the proposed transition matrix, each player has a probability of $1 - \frac{N-1}{N}m$ to remain in the same position, and a probability of $\frac{1}{N}m$ to move to a different one. When $N = 3$, the transition matrix is given by
\[
\bordermatrix{
\text{} & s_{ijk} & s_{ikj} & s_{jik} & s_{jki} & s_{kij} & s_{kji} \cr
s_{ijk} & 1 - \frac{2}{3}m & 0 & 0 & \frac{1}{3}m & \frac{1}{3}m & 0 \cr
s_{ikj} & 0 & 1 - \frac{2}{3}m & \frac{1}{3}m & 0 & 0 & \frac{1}{3}m \cr
s_{jik} & 0 & \frac{1}{3}m & 1 - \frac{2}{3}m & 0 & 0 & \frac{1}{3}m \cr
s_{jki} & \frac{1}{3}m & 0 & 0 & 1 - \frac{2}{3}m & \frac{1}{3}m & 0 \cr
s_{kij} & \frac{1}{3}m & 0 & 0 & \frac{1}{3}m & 1 - \frac{2}{3}m & 0 \cr
s_{kji} & 0 & \frac{1}{3}m & \frac{1}{3}m & 0 & 0 & 1 - \frac{2}{3}m \cr
}
\]
As before, when $m = 0$, players remain in the same position with certainty, while when $m = 1$, they have the same probability to move to any position.\footnote{Our mobility parameter \( m \) can be interpreted analogously to a class of mobility indices, such as the Prais index (\citealp{prais1955}). In our setting, applying the Prais index to the transition matrix yields \( \text{Prais}(P) = \frac{m}{N - 1} \).
This aligns with our interpretation of \( m \) as the degree of income mobility. While we do not propose a new index, our framework embeds this parametric measure of mobility directly into a repeated-game structure. 
}

Despite its simplicity, the transition matrix is in line with the definition of relative social mobility proposed in \cite{behrman2000social}: ``\emph{Holding total income and income distribution constant, after all, relative social mobility is greater if wealthier people more frequently change places with poorer people than if such exchanges occur less frequently}” (p. 74).\footnote{A limitation of our transition matrix is that moving across positions does not depend on the distance between them. For example, in the $N = 3$ case, the players in the second and third income rank have the same probability of moving to the first income rank. Accounting for this would require a more complex transition matrix with additional parameters, complicating the analysis.} 
\subsection{Income inequality} \label{Incomeinequality}
Following the \textit{Atkinson index} \citep{ATKINSON1970244},
we define each income level as follows:

\begin{equation}\label{eq:w}
w_{i} (\alpha )=\frac{\exp (\alpha (N-i+1))}{\sum _{j=1}^{N}\exp (\alpha (N-j+1))} \in [0,1],
\end{equation}
where $i\in \{1,2,\dotsc ,N\}$ is the index of the income type, and $\alpha \geq 0$ is a parameter that governs the degree of inequality. For any $\alpha \geq 0$, the total income is normalized to 1, $\sum _{i=1}^{N} w_{i} (\alpha )=1$.
\par 
When $\alpha = 0$, there is full income equality, with $w_i = \frac{1}{N}$ for any $i \in \{1, ..., N\}$. As $\alpha$ increases, higher income types are allocated progressively larger shares, thereby increasing income inequality. In the limit, as $\alpha \to \infty$, there is full income inequality, with $w_1 = 1$ and $w_i = 0$ for any $i \in \{2, ..., N\}$. Thus, higher values of $\alpha$ are associated with greater income inequality within the organization.

\subsection{Types of sharing rules} \label{Sharing}
We study the sustainability of sharing rules in which players' contributions to the organization may be a function of their income level. 
More concretely, we focus on the parametrization where the share contributed by an player with income $w_i$ is given by:
\begin{equation}
\theta(w_i) = w_i^{\beta} \in [0, 1],
\label{eq:theta_function}
\end{equation}
where $\beta \geq 0$ determines the \textit{degree of progressivity} of the sharing rule.\footnote{Since $w_i$ is between $0$ and $1$, $\theta(w_i)$ is also between $0$ and $1$, consistent with interpreting $\theta(w_i)$ as the share of income one should transfer.} When $\beta = 0$, all players should contribute all their endowment. If $0 < \beta < 1$, the contribution rule is \textit{regressive}: high-income players should contribute more in absolute terms, but less in relative terms. When $\beta = 1$, contributions are exactly proportional to income. When $\beta > 1$, the contribution rule is \textit{progressive}: higher-income players should contribute more than proportionally. 


In our main analysis, we consider \textit{contribution norms} as (exogenous) parameters that societies are endowed with. That is, rather than deriving these norms from individual optimization, we take them as given and explore how income mobility and income inequality affect their sustainability.\footnote{In Section \ref{endogenous}, we endogenize \textit{contribution norms} with evolutionary foundations.}  We motivate this with the findings of \cite{reuben2013enforcement}, who document, in a public goods game experiment with heterogeneous endowments, that income heterogeneity gives rise to ``\emph{a plurality of normatively appealing rules of behavior that are potential candidates for emerging \textit{contribution norm}s}" (p. 15).  
\subsection{Visual representation when $N = 2$} \label{Visual}
Figure 1 displays a visual representation of the game when $N = 2$. In this case, each period has two possible states depending on the income ranks assigned to players $i$ and $j$. At $t = 1$, the game starts in state $s_{i, j}$ or $s_{j, i}$. Players $i$ and $j$ simultaneously choose the share of their income to transfer to the organization. Both players observe the transfers, and their monetary payoffs are determined according to equation \ref{payoff}. At $t = 2$, with probability $1 - \frac{1}{2}m$, the players remain in the same position, and with probability $\frac{1}{2}m$, they switch positions. Players observe the realized state and choose their transfers. This process continues indefinitely. 

\begin{figure}[H]
    \begin{center}
    \includegraphics[width= 0.80\textwidth]{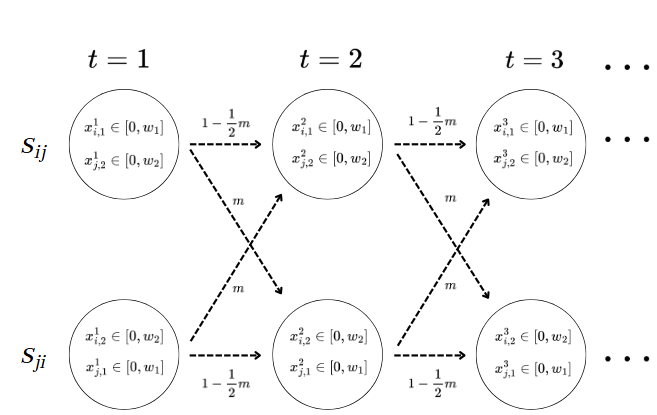}
    \end{center}
    \caption{Visual representation of the $N = 2$ game.}
\end{figure}
\subsection{Notation and Equilibrium Concept}

We now describe the main components of the game when $N = 2$.\footnote{The notation easily generalizes to the case with $N > 2$.} Let $s^t \in \{s_{i,j}, s_{j,i}\}$ denote the state realized in period $t$, and let $s^t = (s^1, s^2, \dots, s^t) \in S^t$ denote the history of states up to period $t$. Correspondingly, let $x_t(s^t)$ denote the profile of contributions made in period $t$ given $s^t$. For instance, if $s^t = s_{i,j}$, then
\begin{equation}
x_t(s_{i,j}) = \{x_{i,1}^t, x_{j,2}^t\},
\end{equation}
where $x_{i,1}^t \in [0, 1]$ is the share of income that player $i$ contributes while occupying the first position. Let $x^t = (x_1(s^1), x_2(s^2), \dots, x_t(s^t))$ denote the full history of observed contributions and states up to period $t$. A \textit{strategy} for player $i$ is a mapping that assigns a contribution level given the full public history $x^{t-1}$ and the current state $s^t$. Formally,
\begin{equation}
x_{i,k}^t : x^{t-1} \times s^t \rightarrow [0,1],
\end{equation}
where $k \in \{1,2\}$ denotes the position occupied by player $i$ in state $s^t$. Players share a common discount factor $\delta \in [0,1)$ and maximize their expected discounted sum of utility over an infinite horizon. The \textit{equilibrium concept} is Subgame Perfect Nash Equilibrium (SPNE). A strategy profile $\{x_i^*, x_j^*\}$ is an SPNE if, at every history $(x^{t-1}, s^t)$, the continuation strategies form a Nash equilibrium of the subgame starting at that point. That is, for each player $i$,
\begin{equation}
x_i^* \in \arg\max_{\hat{x}_i} U_i(\hat{x}_i, x_j^* \mid x^{t-1}, s^t),
\end{equation}
where $U_i$ denotes player $i$'s expected discounted utility from period $t$ onward.

\section{General solution} \label{GeneralSol}
We characterize the conditions under which cooperation can be sustained in all periods. 
We define the \textit{cooperative strategy} as the strategy in which the player follows the \textit{contribution norm} as long as others also do so. For player $i$, the \textit{cooperative strategy} is defined as follows:

\vspace{-8pt}
\begin{itemize}
\itemsep0em 
\item At $t = 1$, select $x_{i}^1 = w_i^{\beta}$.
\item For $t > 1$:  
If in all previous periods $\tau < t$, each player $j \in \{1, \dots, N\}$ has selected $x_{j, k}^{\tau} = w_k^{\beta}$ for their respective position $k$, then choose $x_{i, k}^t = w_k^{\beta}$.  
Otherwise, select $x_{i, k}^t = 0$.
\end{itemize}
\vspace{-8pt}

The \textit{cooperative outcome} corresponds to the outcome in which all players follow the cooperative strategy.
To determine when the cooperative outcome constitutes a SPNE, we compare the discounted expected payoff of cooperation with that of deviation. 
We focus on the most profitable one-shot deviation: contributing zero in the first period.

We denote by \( V_i^c \), \( V_i^d \), and \( V_i^a \), player $i$'s expected discounted payoffs from \textit{cooperation}, \textit{deviation}, and \textit{autarky}, respectively. Under the cooperative outcome, each player contributes according to the rule \( \theta(w_i) = w_i^\beta \) in every period. The value of cooperation is 
\begin{equation}
V_i^c = u(w_i^c) + \delta \left[ \left(1 - \frac{N - 1}{N}m \right) V_i^c 
+ \frac{m}{N} \sum_{j \neq i} V_j^c \right],
\end{equation}
where
\begin{equation} \label{Tra}
w_i^c = [1 - \theta(w_i)] w_i + \frac{1}{N} \sum_{j = 1}^{N} \theta(w_j) w_j.
\end{equation}

The first term of equation \ref{Tra} corresponds to the income kept after contributing \( \theta(w_i) w_i \), while the second term reflects the equally shared benefit from the public good. 
The value of deviation is
\begin{equation} \label{10}
V_i^d = u\left( w_i + \frac{1}{N} \sum_{j \neq i} \theta(w_j) w_j \right) 
+ \delta \left[ \left(1 - \frac{N - 1}{N}m \right) V_i^a 
+ \frac{m}{N} \sum_{j \neq i} V_j^a \right].
\end{equation}
The first term of equation \ref{10} represents the one-shot gain from not contributing, while
the second term represents the discounted continuation value under autarky, in which all players revert permanently to the non-cooperative outcome, defined as:
\begin{equation}
V_i^a = u(w_i) + \delta \left[ \left(1 - \frac{N-1}{N}m \right) V_i^a 
+ \frac{m}{N} \sum_{j \neq i} V_j^a \right].
\end{equation}

\noindent In Appendix \ref{AppA}, we derive closed-form expressions for \( V_i^c \), \( V_i^d \), and $V_i^a$ as functions of the model’s primitives. Lemma~\ref{le:richest_CS} characterizes how the first-type player’s values of cooperation and deviation, $V_1^c$ and $V_1^d$, change with $m$ and $\alpha$. 



\begin{lemma}[Comparative statics for type \( i = 1 \)] \label{le:richest_CS}
Let \( V_1^c \) and \( V_1^d \) denote the cooperative and deviation payoffs for the richest type. Then, there exists a threshold function \( m(\beta) > 0 \) such that:
\begin{align}
\frac{\partial V_1^c}{\partial m} 
  &= 
  \begin{cases}
    0 & \text{if } \beta = 0, \\
    < 0 & \text{if } \beta > 0,
  \end{cases}
&\qquad
\frac{\partial V_1^d}{\partial m} 
  &< 0, \notag \\[1ex]
\frac{\partial V_1^c}{\partial \alpha} 
  &= 
  \begin{cases}
    0 & \text{if } \beta = 0, \\
    < 0 & \text{if } \beta > 0,
  \end{cases}
&\qquad
\frac{\partial V_1^d}{\partial \alpha} 
  &= 
  \begin{cases}
    0 & \text{if } \beta = 0, \\
    < 0 & \text{if } \beta \in (0, m(\beta)), \\
    > 0 & \text{if } \beta > m(\beta).
  \end{cases} \notag
\end{align}
\end{lemma}



Lemma~\ref{le:richest_CS} characterizes how the value of cooperation and the value of deviation for the richest player changes with income mobility (\( m \)) and income inequality (\( \alpha \)). When \( \beta = 0 \), all individuals contribute the same share of their income regardless of their type, and changes in \( m \) and \( \alpha \) do not affect the richest player’s value from cooperation. When \( \beta > 0 \), an increase in mobility reduces both the value of cooperation and the value of deviation, as it raises uncertainty about future positions and weakens the link between present actions and future benefits. The effect of inequality on the deviation value is non-monotonic: for low values of \( \beta \), rising inequality discourages deviation, but when \( \beta \) is high, the burden of contributing rises faster than the corresponding benefit, making deviation more attractive. This trade-off is captured by a threshold function \( m(\beta) \), which separates the two regimes.

\begin{lemma} \label{le:2}
Let $N \geq 2$, $w_1 > w_2 > \dots>w_N$, $\delta \in (0,1)$, $m \in [0,1]$, and $\beta\geq0$. Then, the value of deviation satisfies: $V_{1}^{dev} > V_{2}^{dev} > \dots > V_{N}^{dev}$.
\end{lemma}

Lemma~\ref{le:2} shows that richer individuals have stronger incentives to deviate from cooperation. This holds even when the value of cooperation varies across types due to redistributive norms (i.e., $\beta > 0$), because the private gains from defection and the continuation value in autarky both increase with income and outweigh the non-monotonic effects arising from cooperation under progressive norms. This observation plays a crucial role in determining whether cooperation can be sustained in equilibrium: if the richest type finds it optimal to cooperate, then all lower-income types—who face weaker incentives to defect—will also choose to cooperate. Thus, the sustainability of cooperation boils down to checking the richest' type incentive condition. Proposition~\ref{prop:cooperation} formalizes this insight.

\begin{proposition} \label{prop:cooperation}
The set of discount factors that guarantees cooperation for all individuals is given by the interval $D = [\underline{\delta}, 1]$, where \( \underline{\delta} \in (0,1) \) is the unique discount factor such that $V_1^c = V_1^d$.
\end{proposition}

Proposition~\ref{prop:cooperation} characterizes the conditions under which cooperation can be sustained in equilibrium. The result indicates that the richest individual typically faces the strongest incentive to defect. If this individual prefers to adhere to the cooperative strategy, others will follow as well. The minimum discount factor $\underline{\delta}$ is implicitly defined by the indifference point where the richest player's cooperative and deviation payoffs coincide. Therefore, cooperation can be sustained in equilibrium whenever $\delta \geq \underline{\delta}$.


\section{The effect of income mobility and income inequality} \label{Comparative}
In this section we study how does $\underline{\delta}$ vary with $m$ and $\alpha$.
\subsection{The role of income mobility} \label{ImpactIncome}
Lemma~\ref{lemma:delta_m} shows that increasing income mobility has a positive effect on sustaining cooperation.
\begin{lemma} \label{lemma:delta_m}
Let \( \underline{\delta} \) denote the minimum discount factor that sustains cooperation for all individuals, as defined in Proposition~\ref{prop:cooperation}. Then:
\begin{itemize}
    \item[(i)] If \( m \in (0, 1]  \), then $\underline{\delta} < 1$. If $m = 0$, then $\underline{\delta} = 1$.
    \item[(ii)] For all $m \in (0, 1]$, then  $\frac{\partial \, \underline{\delta}}{\partial m} < 0$.

\end{itemize}
\end{lemma}
When \( m = 0 \), cooperation cannot be sustained for any $\delta < 0$. Intuitively, the richest individual is certain that it will remain forever in the highest position, and therefore its autarky payoff is higher than the cooperation payoff. In this case, high-income players have nothing to lose from deviation, as they enjoy a higher utility outside the cooperative agreement. 
When \( m > 0 \), future income becomes uncertain. This uncertainty reduces the long-run value of deviation because a rich player is more likely to transition into a lower-income position, where the value of autarky is lower. In contrast, the cooperative payoff remains relatively stable due to the smoothing effect of public good provision. Increasing $m$, amplifies the gap between the relatively stable cooperative payoff and the declining deviation payoff, thereby reducing the minimum discount factor $\underline{\delta}$ required to sustain cooperation.\footnote{The effect is particularly strong when \( \beta = 0 \), where contributions are independent of income and cooperative payoffs remain constant across states. Here, mobility reduces the temptation to deviate without diminishing cooperative gains, making cooperation increasingly easier to sustain. }

\begin{figure}[H]
    \centering
    \includegraphics[width=1\linewidth]{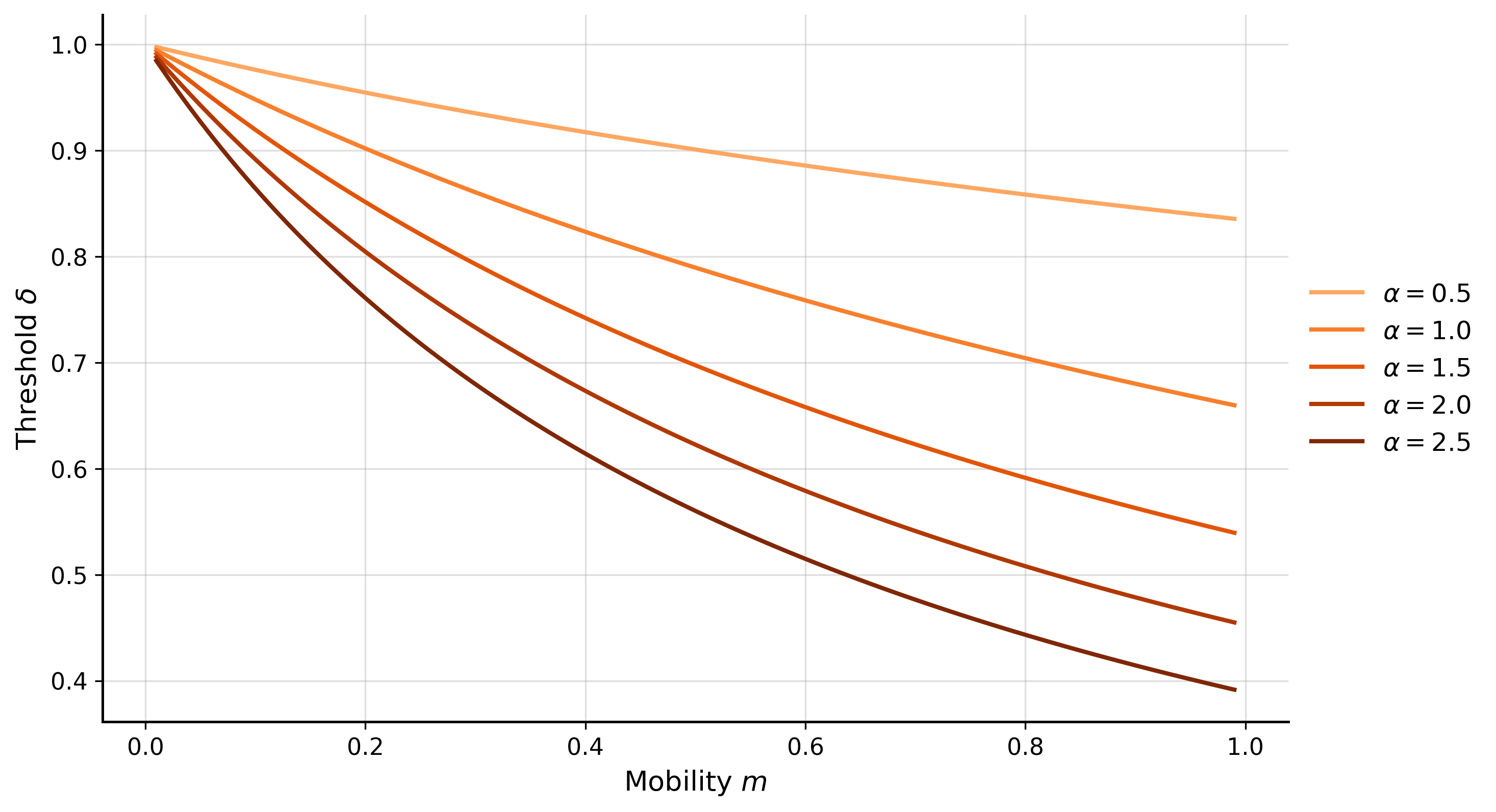}
    \caption{Minimum discount factor $\underline{\delta}$ required for cooperation as a function of income mobility $m$, for different levels of inequality $\alpha$. Parameters: $\rho = 1$, $\beta = 1$, $N = 3$.}
    \label{fig:delta_m}
\end{figure}

Figure~\ref{fig:delta_m} illustrates how $\underline{\delta}$ varies with income mobility $m$, across different levels of inequality $\alpha$ when $\beta = 0$.

\subsection{The role of income inequality}
\label{ImpactInequality}

To study the effect of a permanent increase in inequality in all future periods, we distinguish between inequality at \( t = 0 \) and inequality in future periods \( t > 0 \). To do so, we fix \( \alpha_0 \geq 0 \), and consider an increase to \( \alpha_1 > \alpha_0 \) in all subsequent periods. This allows us to study an increase in future inequality, while keeping total income constant across periods, while abstracting from income effects at $t = 0$. Lemma \ref{le:3} shows that the effect of inequality on cooperation crucially depends on $\beta$ and $m$. 




\begin{lemma} \label{le:3}
Let $ \underline{\delta}(\alpha_1; m, \beta, \alpha_0) $ denote the minimum discount factor required to sustain cooperation when future inequality is $ \alpha_1 $, inequality at $t = 0$ is $\alpha_0 > 0 $, income mobility is $ m \in (0,1] $, and the \textit{contribution norm}’s progressivity is $ \beta \geq 0 $. Then:

\begin{itemize}
    \item[(i)] If $ \beta \leq 1 $, then $ \underline{\delta}(\alpha_1; m, \beta, \alpha_0) $ is weakly decreasing in $ \alpha_1 $.
    
    \item[(ii)] If $ \beta > 1 $, the function $ \underline{\delta}(\alpha_1; m, \beta, \alpha_0) $ is generally non-monotonic with a single-peaked shape.
    
    \item[(iii)] In both cases, the marginal effect of $ \alpha_1 $ on $ \underline{\delta}(\alpha_1; m, \beta, \alpha_0) $ is decreasing in $ m $.

\end{itemize}
\end{lemma}

Lemma~\ref{le:3} shows that the effect of future inequality on the sustainability of cooperation depends crucially on the progressivity of the sharing norm and the degree of income mobility. When $\beta < 1$, the norm is regressive, so high-income individuals are not excessively burdened by contributions. In contrast, when $\beta > 1$, the norm is progressive placing a heavier burden on the rich. As inequality rises, so does their cost of cooperation, increasing the temptation to deviate. This initially raises the cooperation threshold $\underline{\delta}$. However, at high enough levels of inequality, autarky becomes less attractive: income mobility increases the chance of transitioning to a worse position, reducing the expected value of deviation. As a result, the threshold $\underline{\delta}$ eventually declines. 
\par 
This generates a single-peaked relationship between inequality and the cooperation threshold. Importantly, when $\beta > 1$, the sign of the effect of inequality on cooperation depends on the level of mobility: for low $m$, inequality may increase the cooperation threshold (making cooperation harder to sustain), while for high $m$, it lowers the threshold (making cooperation easier to sustain). Finally, (iii) captures the complementarity between inequality and mobility: greater mobility amplifies the stabilizing effect of inequality on cooperation, further lowering the discount factor required to sustain it.

\begin{figure}[H]
    \centering
    \begin{minipage}[c]{0.49\textwidth}
        \centering
        \includegraphics[height=5cm]{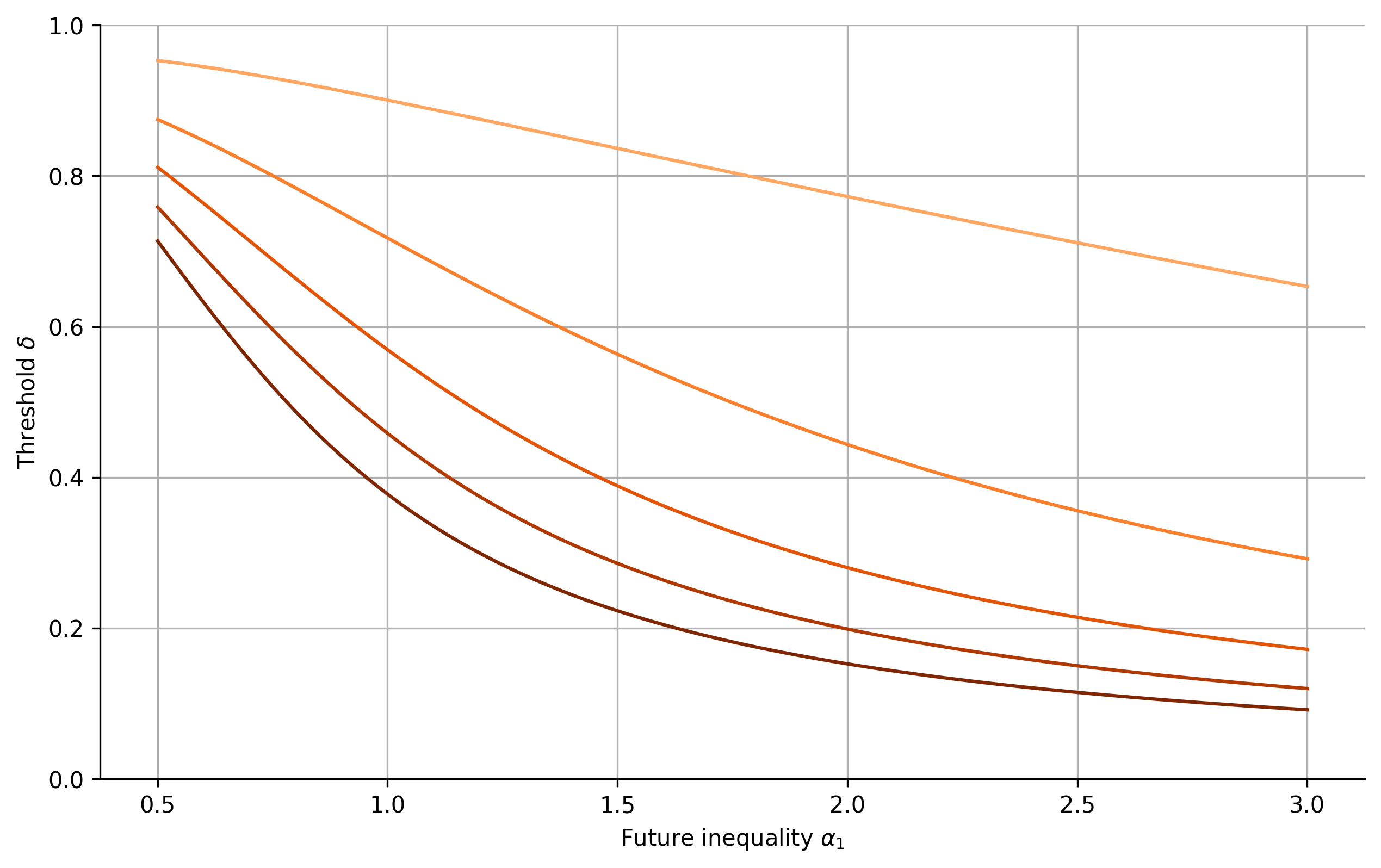}
        \caption*{\small $\beta = 1$ (Proportional contributions)}
    \end{minipage}
    \hfill
    \begin{minipage}[c]{0.49\textwidth}
        \centering
        \includegraphics[height=5cm]{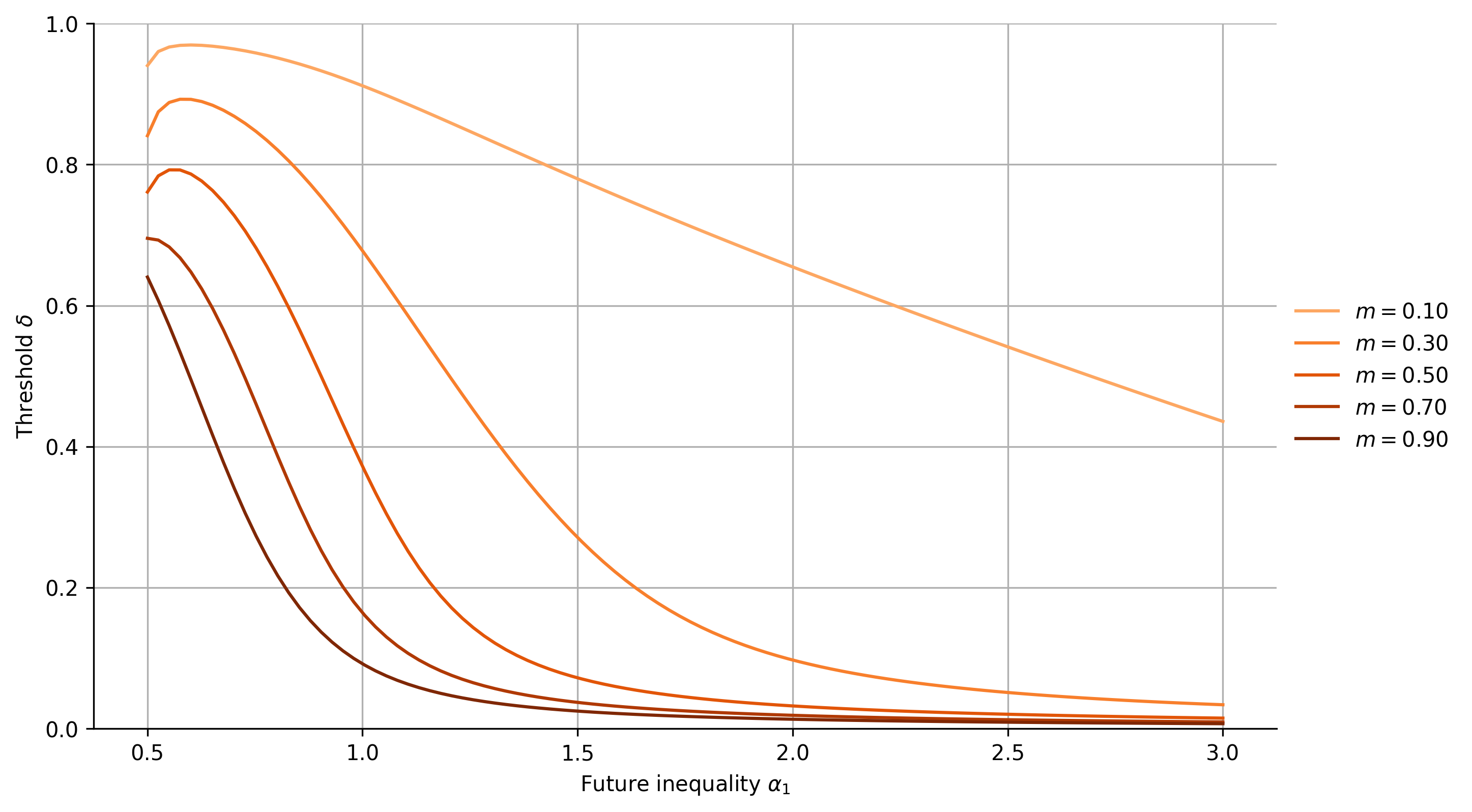}
        \caption*{\small $\beta = 4$ (Progressive contributions)}
    \end{minipage}
    \caption{Minimum discount factor $\underline{\delta}$ required for cooperation as a function of future inequality $\alpha_1$, for various levels of income mobility $m$. 
    Parameters: $\rho = 1$, $\alpha_0 = 0.5$, $N = 5$.}
    \label{fig:delta_alpha1_comparison}
\end{figure}
\subsection{Long-term norm selection} \label{endogenous}

We now study which \textit{contribution norm} $\beta$ is most conducive to the sustainability of cooperation in the long run. Our aim is to endogenize $\beta$ by appealing to an evolutionary logic. Consider a population in which \textit{contribution norms} are subject to selection: groups that fail to sustain cooperation are less likely to survive or be imitated, while groups that succeed in maintaining cooperation are more likely to reproduce their norm.

Formally, let $\underline{\delta}(\beta; \alpha, m)$ denote the minimum discount factor required to sustain cooperation, given inequality $\alpha$, income mobility $m$, and the norm's progressivity $\beta$. Assume individuals have heterogeneous discount factors, with $\delta_i \in (0,1)$ independently drawn from a continuous distribution $F(\cdot)$. For any given $(\alpha, m)$, cooperation is viable for individuals with $\delta_i \geq \underline{\delta}(\beta; \alpha, m)$. Therefore, the share of individuals who can cooperate under a given norm is given by $1 - F\left( \underline{\delta}(\beta; \alpha, m) \right).$ \par
\begin{figure}[H]
    \centering
    \includegraphics[width=0.85\linewidth]{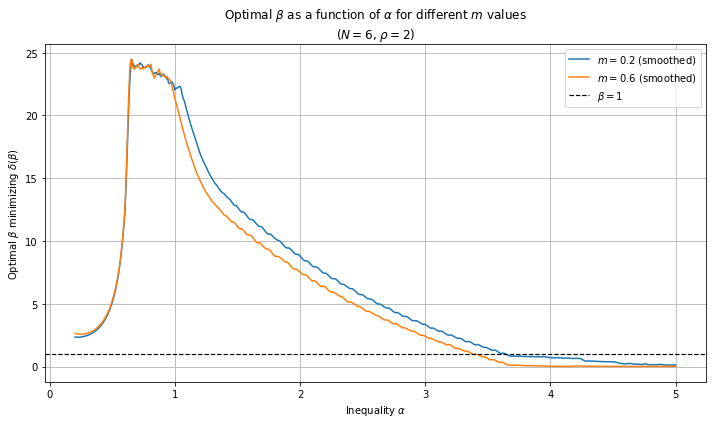}
    \caption{$\beta^*(\alpha, m)$ as a function of $\alpha$, for two values of $m$. Each curve reports the value of $\beta$ that minimizes the cooperation threshold $\underline{\delta}(\beta; \alpha, m)$, computed via a two-stage grid search with local refinement (using 1000 values for $\beta$). For clarity, the resulting series is smoothed using a Savitzky--Golay filter. The dashed line marks $\beta = 1$, corresponding to a flat contribution rule. Parameters: $N = 6$, $\rho = 4$. }
    \label{fig:enter-label}
\end{figure}

 
Selection favor the norm with the progressivity that minimize $\underline{\delta}$, as they maximize the probability of sustaining cooperation within the group. This motivates defining the long-run norm as:
\begin{align}
\beta^*(\alpha, m) = \arg\min_{\beta \geq 0} \, \underline{\delta}(\beta; \alpha, m).
\end{align}
Proposition~\ref{P2} characterizes how $\beta^*(\alpha, m)$ varies with inequality and mobility.\footnote{These properties are established computationally using the method described in Figure~\ref{fig:enter-label}, which numerically minimizes the cooperation threshold over a grid of $\beta$ values.}

\begin{proposition}[Long-run norm selection] \label{P2}
Let $\beta^*(\alpha, m) = \arg\min_{\beta \geq 0} \, \underline{\delta}(\beta; \alpha, m)$. 
Then:
\begin{enumerate}
    \item For fixed $m$, the mapping $\alpha \mapsto \beta^*(\alpha, m)$ is generally non-monotonic: it tends to increase with inequality at low levels of $\alpha$, and decrease at high levels of $\alpha$.
    \item For fixed $\alpha$, higher mobility $m$ is associated with a lower optimal norm $\beta^*(\alpha, m)$.
\end{enumerate}
\end{proposition}

The selected norm $\beta^*(\alpha, m)$ can be either progressive ($\beta^* > 1$) or regressive ($\beta^* < 1$), depending on the environment. When inequality is low, progressive norms are more effective at triggering cooperation by encouraging greater transfers from high earners. As inequality rises, however, overly progressive norms become harder to sustain, and regressive rules may perform better by reducing the burden on top earners. Similarly, higher mobility reduces the need for strong norm enforcement, since individuals are more willing to cooperate in anticipation of future income changes.

Figure~\ref{fig:enter-label} illustrates the relationship between inequality and the optimal \textit{contribution norm} for two mobility values. When inequality is low, the optimal norm rises steeply as progressive rules help enforce cooperation. However, beyond a critical point, further increases in inequality reduce the sustainability of steep norms, and $\beta^*$ declines. The figure also shows that for any level of inequality, higher mobility leads to a lower optimal norm, consistent with Proposition~\ref{P2}.\footnote{The function $\underline{\delta}(\beta; \alpha, m)$ is computed numerically over a discrete grid. As a result, $\beta^*(\alpha, m)$ may exhibit small jumps or flat regions, especially around local minima. Proposition~\ref{P2} focuses on the robust qualitative patterns.}

\section{Application: Government's Optimal Taxation} \label{GovReds}

We extend the baseline model by introducing a preliminary stage in which the government sets a (time-invariant) proportional income tax. This tax is set before the repeated game of voluntary transfers begins. Players then engage in private redistribution on each period, based on their disposable (post-tax) incomes. This two-stage structure allows us to study how government taxation interacts with endogenous cooperation and to what extent public policy can facilitate—or crowd out—informal redistribution.

To analyze the government’s role, we consider a setting in which each player receives a pre-tax income \( w_i \), and the government levies a 
proportional tax schedule:
\begin{align}
T(w_i) = \tau w_i,
\end{align}
where \( \tau \in [0,1] \) is the tax rate.
Post-tax consumption, after voluntary transfers and public redistribution, is given by:
\begin{equation} \label{15}
x_i(\theta) = y_i [1 - \theta(y_i)] + \frac{1}{N} \sum_j \theta(y_j) y_j + \frac{1 - s}{N} \underbrace{\sum_j T(w_j)}_{= \ \tau}.
\end{equation}
The first term of equation \ref{15} represents what individual $i$ keeps after making their voluntary contribution, the second term reflects their equal share of the collectively provided transfers, and the third term captures their share of redistributed tax revenue. The parameter $s \in (0,1]$ captures the exogenous cost of public funds, which could stem from administrative burdens or distortionary effects of taxation.

When $s > 0$, public redistribution is less efficient than private redistribution, which creates a trade-off for the planner. If $s = 0$, redistribution through taxes would dominate, making voluntary transfers irrelevant and leading the government to fully equalize post-tax incomes by setting $T(w_i) = w_i$. To maintain a meaningful role for voluntary transfers, we assume $s > 0$ throughout.\footnote{This is akin to assuming that the government faces an exogenous revenue requirement $R > 0$, leading to a constraint of the form $G^P + R \leq T$ (\citealp{Sandmo1975}).}


\subsection{The Planner's Problem} \label{Welfare}

The planner's objective is to maximize social welfare by choosing an optimal redistribution policy $ \tau $. Welfare depends not only on the resulting payoffs under autarky and cooperation, but also on whether cooperation is sustainable given the underlying incentives. Formally, the planner evaluates:
\begin{equation} \label{welfare}
W(\tau;\delta) =
\begin{cases}
\hat{V}_i^a(\tau) & \text{if } \delta < \underline{\delta}(\tau), \\\\
\hat{V}_i^c(\tau) & \text{if } \delta \geq \underline{\delta}(\tau),
\end{cases}
\end{equation}
where $ \underline{\delta}(\tau) $ denotes the minimum discount factor that makes cooperation incentive-compatible under tax policy $ \tau $. The functions $ \hat{V}_i^a(\tau) $ and $ \hat{V}_i^c(\tau) $ represent average utility across players in the autarkic and cooperative outcomes, respectively:
\begin{align}
\hat{V}_i^a(\tau) &= \frac{1}{N} \sum_{i=1}^{N} u\left(x_i^{a}(\tau)\right), \\
\hat{V}_i^c(\tau) &= \frac{1}{N} \sum_{i=1}^{N} u_i^{c}(\tau),
\end{align}
where $ x_i^a(\tau) $ denotes individual consumption under autarky. The planner chooses the tax rate $\tau$ to maximize expected welfare, taking into account both economic outcomes and the incentives for cooperation.
A central challenge is that both the value of cooperation and the incentive constraint—summarized by $\underline{\delta}(\tau)$—depend on the tax policy itself. Even when the discount factor $\delta$ is fixed, the planner may be able to induce or discourage cooperation through their choice of $\tau \in [0, 1]$. Taxation thus shapes not only the redistribution of income, but also the sustainability of cooperative behavior in the repeated game.

Proposition~\ref{prop:tax} formalizes the distinction between two regimes: when cooperation can be sustained and when it cannot. It also describes how the optimal tax responds to changes in inequality, mobility, and norm's progressivity in each regime.

\begin{proposition} \label{prop:tax}
Let \( \underline{\delta}(\tau) \) denote the minimal discount factor required to sustain cooperation at tax rate \( \tau \).  
Let \( \tau^a(\alpha) = \arg\max_{\tau \in [0,1]} \hat{V}_i^a(\tau) \) denote the optimal tax rate under autarky.

Then:
\begin{itemize}
    \item If there exists \( \tau \in [0,1] \) such that \( \delta = \underline{\delta}(\tau) \), then cooperation is attainable, and the planner implements:
    \[
    \tau^* = \tau^\dagger(\delta) \quad \text{such that} \quad \underline{\delta}(\tau^\dagger) = \delta,
    \]
    with the following comparative statics:
    \[
    \frac{\partial \tau^\dagger}{\partial m} < 0, \quad
    \frac{\partial \tau^\dagger}{\partial \beta} < 0, \quad
    \frac{\partial \tau^\dagger}{\partial \alpha} < 0.
    \]

    \item If no such \( \tau \in [0,1] \) exists (i.e., \( \delta < \min_{\tau} \underline{\delta}(\tau) \)), then cooperation is not sustainable, and the planner chooses:
    \[
    \tau^* = \tau^a(\alpha),
    \]
    with:
    \[
    \frac{\partial \tau^a}{\partial m} = 0, \quad
    \frac{\partial \tau^a}{\partial \beta} = 0, \quad
    \frac{\partial \tau^a}{\partial \alpha} > 0.
    \]
\end{itemize}
\end{proposition}

Proposition~\ref{prop:tax} distinguishes two regimes depending on whether cooperation can be sustained at any feasible tax rate. In the autarkic regime, where \( \delta < \underline{\delta}(\tau) \) for all \( \tau \in [0,1] \), private cooperation is not feasible. Therefore, redistribution occurs solely through taxation. In this case, the optimal tax rate \( \tau^a \) depends only on inequality: as \( \alpha \) increases, the planner’s incentive to reduce income dispersion leads to a higher tax. Since private transfers are absent in this regime, income mobility \( m \) and the norm's progressivity \( \beta \) have no effect on the planner’s choice.

In the cooperative regime, where \( \delta = \underline{\delta}(\tau) \) for some \( \tau \in [0,1] \), the planner implements cooperation exactly at the threshold of sustainability. That is, the planner selects a tax rate \( \tau^\dagger(\delta) \) such that cooperation becomes just viable in equilibrium. This tax rate enables both public and private redistribution. Because private transfers depend on income mobility and the norm's progressivity, higher \( m \) and \( \beta \) improve the effectiveness of voluntary redistribution, allowing the planner to reduce reliance on taxation. Moreover, greater inequality \( \alpha \) increases the volume of private transfers once cooperation is in place, further reducing the optimal tax burden. As a result, the optimal cooperative tax rate \( \tau^\dagger \) is decreasing in all three parameters: \( m \), \( \beta \), and \( \alpha \).

\begin{figure}[H]
    \centering
    \includegraphics[width=0.8\linewidth]{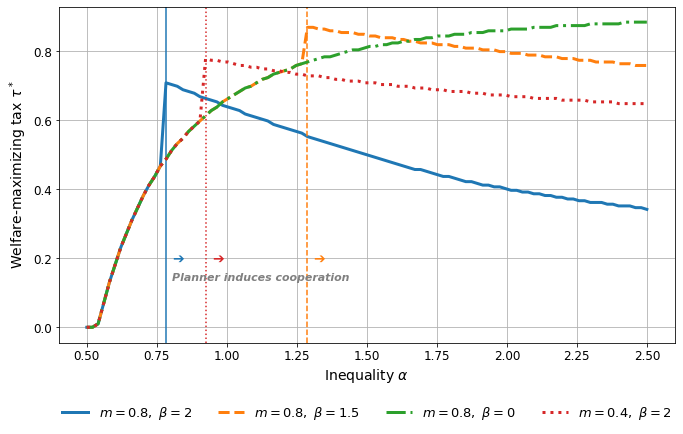}
    \caption{ Welfare-maximizing tax rate $\tau^*$ as a function of inequality $\alpha$, for different values of income mobility $m$ and norm's progressivity $\beta$.  
    Parameters: $\rho = 0.5$, $\mu = 1$, $s = 0.09$, $\delta = 0.7$, $N = 3$. 
    }
    \label{fig:tau_star_vs_alpha}
\end{figure}

Figure~\ref{fig:tau_star_vs_alpha} shows how the welfare-maximizing tax rate $\tau^*$ varies with inequality $\alpha$ for four combinations of income mobility $m$ and norm's progressivity $\beta$. Each curve exhibits a threshold value $\alpha$ at which cooperation becomes sustainable. To the left of this threshold, the planner operates in the autarkic regime; to the right, in the cooperative regime. The transition is marked by a vertical line for each curve. 

Comparing the blue and orange lines (both with $m = 0.8$), we see that increasing the norm's progressivity from $\beta = 1.5$ to $\beta = 2$ shifts the cooperation threshold leftward and lowers the optimal tax under cooperation. This illustrates that more progressive norms facilitate cooperation at lower inequality levels and reduce the need for taxation. Next, comparing the blue and red lines (both with $\beta = 2$), we observe that reducing mobility from $m = 0.8$ to $m = 0.4$ enables cooperation to emerge at a lower level of inequality. It also leads to a higher cooperative tax rate, since lower mobility reduces endogenous transfers. The green line, with $m = 0.8$ and $\beta = 0$, serves as a benchmark without public cooperation. This trajectory aligns with the autarkic segments of the other curves, confirming that in the absence of cooperation, the optimal tax rises with inequality and is unaffected by $m$ or $\beta$.

\section{Conclusions} \label{Conclusions}
This paper studies how income mobility and income inequality jointly shape the sustainability of cooperation in repeated interactions. We introduce a tractable framework of a risk-sharing organization in which individuals decide how much of their income to transfer to others. A key feature of the model is that individuals change their income over time, while keeping the organization’s income distribution constant. We characterize the conditions under which \textit{\textit{contribution norm}s}—informal agreements about how much each member should transfer—can be sustained in equilibrium. Specifically, we find that income mobility has a positive effect on sustaining cooperation. In contrast, the effect of inequality is ambiguous and depends on the progressivity of the \textit{contribution norm} and the degree of mobility. 

Our framework relates to a variety of contexts where cooperation is crucial. Examples include risk-sharing between workers (\citealp{ligon2002informal}; \citealp{bigsten2003risk}), local tax agreements (\citealp{wolf2010community}), village-based health insurance schemes (\citealp{jajoo1992risk}; \citealp{ahmed2016willingness}), or fiscal arrangements within regions or monetary unions (\citealp{ventura2019joseph}). In all these cases, participants repeatedly interact, face unequal incomes, and must rely on informal or partially enforced norms to govern transfers. Our results suggest that income mobility—whether driven by institutional design or economic shocks—plays a central role in determining the sustainability of such arrangements.  

One could extend the proposed framework in several dimensions. First, while our baseline model assumes self-interested individuals, incorporating social preferences such as altruism (\citealp{becker1974theory}), reciprocity (\citealp{rabin1993incorporating}), or inequity aversion (\citealp{fehr1999theory}) may alter our conclusions. Second, introducing heterogeneity in, for example, risk preferences or income risk could generate additional insights into the effect of income dynamics on cooperation (\citealp{briys1990risk}; \citealp{huber2022comparative}). Third, allowing individuals to save across periods would introduce a trade-off between informal insurance and self-insurance, providing a connection with models of precautionary savings (\citealp{kocherlakota2004figuring}; \citealp{buera2011self}). We leave these extensions for future research.

\newpage
\bibliographystyle{chicago}
\bibliography{sample}
\newpage 
\appendix
\section{Appendix A: Mathematical Proofs} \label{AppA}
\subsection{Preliminaries}
\begin{lemma} \label{le:1}
Let $\overline{u}^{\,aut} = \frac{1}{N} \sum_{j=1}^N u(w_j)$, $u_i^c = u\left( (1 - \theta(w_i)) w_i + \frac{1}{N} \sum_{j=1}^N \theta(w_j) w_j \right)$, and $\overline{u}^{\,coop} = \frac{1}{N} \sum_{j=1}^N u_j^c$. The values of autarky, deviation, and cooperation are given by: 
\begin{align}
V_i^{a} &= \frac{1}{1 - \delta (1 - m)}u(w_i) 
+ \frac{\delta m}{(1 - \delta)(1 - \delta (1 - m))} \, \overline{u}^{\,aut},
\end{align}
\begin{align}
V_i^{d} &= u\left( w_i + \frac{1}{N} \sum_{j \neq i} \theta(w_j) w_j \right) 
+ \delta \left[ (1 - m) \left( \frac{u(w_i) + \delta m \, \overline{u}^{\,aut}}{1 - \delta (1 - m)} \right) 
+ m \, \overline{u}^{\,aut} \right],
\end{align}
\begin{align}
V_i^{c} &= \frac{(1 - \delta) u_i^c + \delta m \, \overline{u}^{\,coop}}{(1 - \delta)(1 - \delta (1 - m))}.
\end{align}
\end{lemma}
\subsection*{Proof of Lemma \ref{le:1}}
\paragraph{Autarky-algebra.} Define the sum of autarky values across income-types as $S=\sum_{j=1}^NV_j^{a}$, we can the re express the equation above as:
\begin{equation}\label{eq:Spre}
   V_i^{a} = u(w_i) + \delta \left[ \left(1 - \frac{N-1}{N}m\right) V_i^{a} + \frac{m}{N} \left( S-V_i^{a}\right) \right].
\end{equation}
If we take the sum of the expression above we obtain:
\begin{equation}
   S= \sum_{i=1}^Nu(w_i) + \delta \left[ \left(1 - \frac{N-1}{N}m\right) S+ \frac{m}{N} \left( S\cdot N-S\right) \right].
\end{equation}
By simple algebra we obtain:
\begin{align}\label{eq:S}
    S=\frac{\sum_{i=1}^Nu(w_i)}{1-\delta}.
\end{align}
Substituting the value of $S$ into the expression for $V_i^{a}$ yields:
\begin{align}\label{eq:au}
    V_i^{a} = \frac{1}{1-\delta(1-m)} \left[ u(w_i)+\frac{\delta m}{N}\frac{\sum_{i=1}^N u(w_i)}{1-\delta}\right].
\end{align}
Autarky values are monotonically increasing in income levels $w_i$.

\paragraph{Algebra of deviations.} We have that:
\begin{align*}
       V_i^{d}&=  u\left( w_i + \frac{\sum_{j\neq i}w_j} {N} \right) + \delta \left[ \left(1 - \frac{N-1}{N}m\right) V_i^{a} + \sum_{j \neq i} \frac{1}{N}m V_j^{a} \right]\\
    &=  u\left( w_i + \frac{\sum_{j\neq i}w_j} {N} \right) + \delta \left[ \left(1 - m\right) \left(\frac{u(w_i) + {\delta} \frac{m}{N} S}{1 - \delta (1 - m)}\right) + \frac{m}{N}S\right]\\
\end{align*}
\subsection*{Proof of Lemma \ref{le:2}}
\begin{proof}
Take $k>i$, then:
   \begin{align*}
   V_i^{d}-V_k^{d} &= u\left( w_i + \frac{\sum_{j\neq i}w_j} {N} \right) -u\left( w_k + \frac{\sum_{j\neq k}w_j} {N} \right)\\
   &+ \delta \left[ \left(1 - \frac{N-1}{N}m\right) \left(V_i^{a}-V_k^{a}\right) + \frac{m}{N}\left(V_k^{a}-V_i^{a}\right) \right]\\ 
   &= u\left( w_i + \frac{\sum_{j\neq i}w_j} {N} \right) -u\left( w_k + \frac{\sum_{j\neq k}w_j} {N} \right)+ \delta \left[ \left(1 - m\right) \left(V_i^{a}-V_k^{a}\right)\right]\\ 
   &>0
\end{align*}
Since both terms in the last line are positive. The instantaneous utility terms decreases in $w_i$, which can be seen by noting that:
\begin{align*}
    w_i + \frac{\sum_{j\neq i}w_j} {N}=w_i\left(\frac{N-1}{N}\right)+\frac{W} {N}.
\end{align*}
for $W=\sum_{j=1}^N w_j$.
\end{proof}

\subsection*{Proof of Lemma \ref{le:3}}
\begin{proof}
    Consider now:
\begin{align*}
    \Delta_i(\delta)&={V_i}^{dev} - {V_i}^{coop}\\
    &=u\left( w_i + \frac{\sum_{j\neq i}w_j} {N} \right) + \delta \left[ \left(1 - m\right) \left(\frac{u(w_i) + {\delta} \frac{m}{N} S}{1 - \delta (1 - m)}\right) + \frac{m}{N}S\right]-\frac{u(\bar{w})}{1-\delta},\\
     &=u\left( w_i + \frac{\sum_{j\neq i}w_j} {N} \right) + \delta \left[ \left(1 - m\right) \left(\frac{u(w_i) + {\delta} \frac{m}{N} \frac{\sum_{i=1}^Nu(w_i)}{1-\delta}}{1 - \delta (1 - m)}\right) + \frac{m}{N}\frac{\sum_{i=1}^Nu(w_i)}{1-\delta}\right]-\frac{u(\bar{w})}{1-\delta},
\end{align*}
where $\bar{w}=\sum_{i=1}^Nw_i/N$ is the average income. We are looking for conditions such that $ \Delta_i(\delta)<0$, which are given by:
\begin{align*}
0& >u\left( w_i + \frac{\sum_{j\neq i}w_j} {N} \right) + \delta \left[ \left(1 - m\right) \left(\frac{u(w_i) + {\delta} \frac{m}{N} \frac{\sum_{i=1}^Nu(w_i)}{1-\delta}}{1 - \delta (1 - m)}\right) + \frac{m}{N}\frac{\sum_{i=1}^Nu(w_i)}{1-\delta}\right] -\frac{u(\bar{w})}{1-\delta}.\\
\end{align*}
\end{proof}
\subsection*{Proof of Proposition \ref{prop:cooperation}}
\begin{proof}
    
Define the coefficients:
\begin{gather*}
c_{0} = u\left( w_{i} + \frac{\sum_{j \neq i} w_{j}}{N} \right), \\
c_{1} = u(w_{i}), \\
c_{2} = u\left(\overline{w}\right) = u\left(\frac{\sum_{i=1}^{N} w_{i}}{N}\right), \\
c_{3} = \frac{1}{N} \sum_{i=1}^{N} u(w_{i}).
\end{gather*}

\noindent The value of deviation is:
\begin{align*}
V_{i}^{\text{dev}} & = u\left(w_{i} + \frac{\sum_{j \neq i} w_{j}}{N}\right) + \delta \left[(1-m)\left(\frac{u(w_{i}) + \delta \frac{m}{N} S}{1-\delta(1-m)}\right) + \frac{m}{N} S\right] \\
& = c_{0} + \delta \left[(1-m)\left(\frac{c_{1} + \delta \frac{m}{1-\delta} c_{3}}{1-\delta(1-m)}\right) + \frac{m}{1-\delta} c_{3}\right].
\end{align*}

\noindent The value of cooperation is:
\begin{align*}
V^{\text{coop}} = \frac{1}{1-\delta}u\left(\frac{\sum_{i=1}^{N} w_{i}}{N}\right) = \frac{1}{1-\delta}c_{2}.
\end{align*}
We aim to analyze:
\begin{align*}
\Delta(\delta) = V^{\text{coop}} - V_{i}^{\text{dev}} > 0,
\end{align*}
which leads to:
\begin{align*}
\Delta(\delta) & = \frac{c_{2}}{1-\delta} - c_{0} - \delta \left[(1-m)\left(\frac{c_{1} + \delta \frac{m}{1-\delta} c_{3}}{1-\delta(1-m)}\right) + \frac{m}{1-\delta} c_{3}\right] > 0. \\
& \Rightarrow c_{2} - c_{0}(1-\delta) - \delta(1-\delta)\left[(1-m)\left(\frac{c_{1} + \delta \frac{m}{1-\delta} c_{3}}{1-\delta(1-m)}\right) + \frac{m}{1-\delta} c_{3}\right] > 0. \\
& \Rightarrow (1-\delta(1-m))c_{2} - (1-\delta(1-m))c_{0}(1-\delta) \\
& \quad - \delta(1-m)\left[(1-\delta)(c_{1} + \delta \frac{m}{1-\delta} c_{3}) + mc_{3}(1-\delta(1-m))\right] > 0. \\
& \Rightarrow c_{2} - c_{0} - \delta((1-m)c_{2} - c_{0} - c_{0}(1-m)) - \delta^{2} c_{0}(1-m) \\
& \quad - \delta((1-m)(c_{1} + \delta(mc_{3} - c_{1})) + mc_{3} - \delta mc_{3}(1-m)) > 0. \\
& \Rightarrow c_{2} - c_{0} + \delta[c_{0} - mc_{3} - (1-m)(c_{2} - c_{0} + c_{1})] \\
& \quad + \delta^{2}(1-m)(c_{1} - c_{0}) > 0.
\end{align*}

\paragraph{Analyzing the roots.}

The quadratic inequality is given by:

\[
c_{2} - c_{0} + \delta \left[ c_{0} - mc_{3} - (1-m)(c_{2} - c_{0} + c_{1}) \right] + \delta^2 (1-m)(c_{1} - c_{0}) > 0.
\]

Rewriting this in standard quadratic form, we have:

\[
A \delta^2 + B \delta + C = 0,
\]

where the coefficients are defined as:

\[
A = (1-m)(c_{1} - c_{0}),
\]
\[
B = c_{0} - mc_{3} - (1-m)(c_{2} - c_{0} + c_{1}),
\]
\[
C = c_{2} - c_{0}.
\]

To analyze the sign of each coefficient, note that from the given inequalities \(c_0 > c_1 > c_2 > c_3\):

1. For \(A\), since \(c_1 - c_0 < 0\) and \(1-m > 0\) for \(m \in [0, 1)\), it follows that \(A = (1-m)(c_1 - c_0) < 0\). Thus, the parabola opens downward.
2. For \(C\), since \(c_2 - c_0 < 0\), it follows that \(C < 0\).
3. For \(B\), we observe that \(c_0 - mc_3 > 0\) (since \(c_0 > c_3\) and \(m \in [0, 1)\)). Additionally, the term \(-(1-m)(c_2 - c_0 + c_1)\) contributes negatively. Therefore, \(B > 0\).

The roots of the quadratic equation are given by:

\[
\delta_{1,2} = \frac{-B \pm \sqrt{B^2 - 4AC}}{2A}.
\]
Consider the negative root $ \delta_{1}$. We can use the fact that $c_0>c_1>c_2>c_4$, and that the discriminant is minimal for either $A=0$ or $C=0$ to find a lower bound for this root:
\begin{align}
    \delta_{1} &\geq \frac{-B + \sqrt{B^2 -0}}{2A}\\
    &=\frac{-B}{A}\\
      &=\frac{c_{0} - mc_{3} - (1-m)(c_{2} - c_{0} + c_{1})}{-(1-m)(c_{1} - c_{0})}\\
        &=\frac{c_{0} - c_{3} }{(c_{1} - c_{0})} +\frac{c_{0} - c_{1} - c_{2} }{(c_{1} - c_{0})}+\frac{mc_{0} + c_{3}-2mc_3 }{(1-m)(c_{1} - c_{0})}\\
        &>1
\end{align}
This means that we only need to take of the second root of the polynomial.\par

\paragraph{Bounds for $\delta_2$.-} Analogously as we did before, the maximal value of the discriminant, obtained by evaluating $A=0$ or $C=0$, yield a lower zero bound for the second root: $\delta_2\geq 0$. Finally, the root is maximal when then discriminant is null, i.e:
\begin{align*}
    \delta_{1} &\leq \frac{-B + \sqrt{0}}{2A}\\
    &=\frac{-B}{2A}\\
     &=\frac{c_{0} - c_{3} }{2(c_{1} - c_{0})} +\frac{c_{0} - c_{1} - c_{2} }{2(c_{1} - c_{0})}+\frac{mc_{0} + c_{3}-2mc_3 }{2(1-m)(c_{1} - c_{0})}\\
     &<1\\
\end{align*}
\end{proof}

\end{document}